%% file: main.tex
\pdfoutput=1

\documentclass[12pt,a4paper]{article}

\usepackage{ifthen} 
\usepackage{graphicx}  
\usepackage{epstopdf}
\newboolean{pdflatex}
\setboolean{pdflatex}{true} 

\newboolean{articletitles}
\setboolean{articletitles}{true} 

\newboolean{uprightparticles}
\setboolean{uprightparticles}{false} 

\input{preamble}
\begin{document}

\renewcommand{\thefootnote}{\fnsymbol{footnote}}
\setcounter{footnote}{1}

\input{title-LHCb-PAPER}

\renewcommand{\thefootnote}{\arabic{footnote}}
\setcounter{footnote}{0}


\pagestyle{plain} 
\setcounter{page}{1}
\pagenumbering{arabic}

\input{paper_v6.tex}

\input{acknowledgements}

\addcontentsline{toc}{section}{References}
\bibliographystyle{LHCb}
\bibliography{main}

\end{document}

%% file: preamble.tex
\usepackage{booktabs} 
\usepackage{microtype}
\usepackage{lineno}  
\usepackage{xspace} 

\usepackage{graphicx}  
\usepackage{color}
\usepackage{colortbl}
\graphicspath{{./figs/}} 

\usepackage{amsmath} 
\usepackage{amssymb}
\usepackage{amsfonts}
\usepackage{upgreek} 

\newcommand*\patchAmsMathEnvironmentForLineno[1]{%
\expandafter\let\csname old#1\expandafter\endcsname\csname #1\endcsname
\expandafter\let\csname oldend#1\expandafter\endcsname\csname
end#1\endcsname
 \renewenvironment{#1}%
   {\linenomath\csname old#1\endcsname}%
   {\csname oldend#1\endcsname\endlinenomath}%
}
\newcommand*\patchBothAmsMathEnvironmentsForLineno[1]{%
  \patchAmsMathEnvironmentForLineno{#1}%
  \patchAmsMathEnvironmentForLineno{#1*}%
}
\AtBeginDocument{%
\patchBothAmsMathEnvironmentsForLineno{equation}%
\patchBothAmsMathEnvironmentsForLineno{align}%
\patchBothAmsMathEnvironmentsForLineno{flalign}%
\patchBothAmsMathEnvironmentsForLineno{alignat}%
\patchBothAmsMathEnvironmentsForLineno{gather}%
\patchBothAmsMathEnvironmentsForLineno{multline}%
}

\usepackage{hyperref}    
\usepackage[all]{hypcap} 

\input{lhcb-symbols-def} 

\usepackage{cite} 
\usepackage{mciteplus}

%% file: lhcb-symbols-def.tex
\def\lhcb {\mbox{LHCb}\xspace}
\def\ux85 {\mbox{UX85}\xspace}

\def\velo   {VELO\xspace}

\ifthenelse{\boolean{uprightparticles}}%
{

 \def\Pmu         {\ensuremath{\upmu}\xspace}

 \def\Ppsi        {\ensuremath{\uppsi}\xspace}

 \def\PDelta      {\ensuremath{\Delta}\xspace}                 
 \def\PXi      {\ensuremath{\Xi}\xspace}                 
 \def\PLambda      {\ensuremath{\Lambda}\xspace}                 
 \def\PSigma      {\ensuremath{\Sigma}\xspace}                 
 \def\POmega      {\ensuremath{\Omega}\xspace}                 
 \def\PUpsilon      {\ensuremath{\Upsilon}\xspace}                 
 
 \def\PB      {\ensuremath{\mathrm{B}}\xspace}                 
                  
 \def\PD      {\ensuremath{\mathrm{D}}\xspace}

 \def\PJ      {\ensuremath{\mathrm{J}}\xspace}                 
 \def\PK      {\ensuremath{\mathrm{K}}\xspace}

 \def\Pb      {\ensuremath{\mathrm{b}}\xspace}                 
 \def\Pc      {\ensuremath{\mathrm{c}}\xspace}

 \def\Pi      {\ensuremath{\mathrm{i}}\xspace}

}
{

 \def\Pmu         {\ensuremath{\mu}\xspace}

 \def\Ppsi        {\ensuremath{\psi}\xspace}                 
                  
 \mathchardef\PDelta="7101
 \mathchardef\PXi="7104
 \mathchardef\PLambda="7103
 \mathchardef\PSigma="7106
 \mathchardef\POmega="710A
 \mathchardef\PUpsilon="7107
                  
 \def\PB      {\ensuremath{B}\xspace}                 
                  
 \def\PD      {\ensuremath{D}\xspace}

 \def\PJ      {\ensuremath{J}\xspace}                 
 \def\PK      {\ensuremath{K}\xspace}

 \def\Pb      {\ensuremath{b}\xspace}                 
 \def\Pc      {\ensuremath{c}\xspace}

 \def\Pi      {\ensuremath{i}\xspace}

}

\def\mumu       {\ensuremath{\Pmu^+\Pmu^-}\xspace}

\def\cquark    {\ensuremath{\Pc}\xspace}

\def\bquark    {\ensuremath{\Pb}\xspace}
\def\bquarkbar {\ensuremath{\overline \bquark}\xspace}

\def\kaon  {\ensuremath{\PK}\xspace}
  \def\Kbar  {\kern 0.2em\overline{\kern -0.2em \PK}{}\xspace}

\def\Kz    {\ensuremath{\kaon^0}\xspace}
\def\Kzb   {\ensuremath{\Kbar^0}\xspace}
\def\KzKzb {\ensuremath{\Kz \kern -0.16em \Kzb}\xspace}
\def\Kp    {\ensuremath{\kaon^+}\xspace}
\def\Km    {\ensuremath{\kaon^-}\xspace}

\def\KpKm  {\ensuremath{\Kp \kern -0.16em \Km}\xspace}

  \def\Dbar    {\kern 0.2em\overline{\kern -0.2em \PD}{}\xspace}
\def\D       {\ensuremath{\PD}\xspace}

\def\Dz      {\ensuremath{\D^0}\xspace}
\def\Dzb     {\ensuremath{\Dbar^0}\xspace}
\def\DzDzb   {\ensuremath{\Dz {\kern -0.16em \Dzb}}\xspace}
\def\Dp      {\ensuremath{\D^+}\xspace}
\def\Dm      {\ensuremath{\D^-}\xspace}

\def\DpDm    {\ensuremath{\Dp {\kern -0.16em \Dm}}\xspace}

\def\Bbar    {\ensuremath{\kern 0.18em\overline{\kern -0.18em \PB}{}}\xspace}

\def\jpsi     {\ensuremath{{\PJ\mskip -3mu/\mskip -2mu\Ppsi\mskip 2mu}}\xspace}

  \def\Y#1S{\ensuremath{\PUpsilon{(#1S)}}\xspace}

\def\Lbar {\ensuremath{\kern 0.1em\overline{\kern -0.1em\PLambda}}\xspace}

\def\to                 {\ensuremath{\rightarrow}\xspace}

\def\AT#1     {\ensuremath{A_{\mathrm{T}}^{#1}}\xspace}           

\def\C#1      {\ensuremath{\mathcal{C}_{#1}}\xspace}                       
\def\Cp#1     {\ensuremath{\mathcal{C}_{#1}^{'}}\xspace}                    
\def\Ceff#1   {\ensuremath{\mathcal{C}_{#1}^{\mathrm{(eff)}}}\xspace}        
\def\Cpeff#1  {\ensuremath{\mathcal{C}_{#1}^{'\mathrm{(eff)}}}\xspace}       
\def\Ope#1    {\ensuremath{\mathcal{O}_{#1}}\xspace}                       
\def\Opep#1   {\ensuremath{\mathcal{O}_{#1}^{'}}\xspace}                    

\newcommand{\tev}{\ensuremath{\mathrm{\,Te\kern -0.1em V}}\xspace}
\newcommand{\gev}{\ensuremath{\mathrm{\,Ge\kern -0.1em V}}\xspace}
\newcommand{\mev}{\ensuremath{\mathrm{\,Me\kern -0.1em V}}\xspace}
\newcommand{\kev}{\ensuremath{\mathrm{\,ke\kern -0.1em V}}\xspace}
\newcommand{\ev}{\ensuremath{\mathrm{\,e\kern -0.1em V}}\xspace}
\newcommand{\gevc}{\ensuremath{{\mathrm{\,Ge\kern -0.1em V\!/}c}}\xspace}
\newcommand{\mevc}{\ensuremath{{\mathrm{\,Me\kern -0.1em V\!/}c}}\xspace}
\newcommand{\gevcc}{\ensuremath{{\mathrm{\,Ge\kern -0.1em V\!/}c^2}}\xspace}
\newcommand{\gevgevcccc}{\ensuremath{{\mathrm{\,Ge\kern -0.1em V^2\!/}c^4}}\xspace}
\newcommand{\mevcc}{\ensuremath{{\mathrm{\,Me\kern -0.1em V\!/}c^2}}\xspace}

\def\mub{\ensuremath{\rm \,\upmu b}\xspace}

\def\nb {\ensuremath{\rm \,nb}\xspace}
\def\invnb {\ensuremath{\mbox{\,nb}^{-1}}\xspace}

\def\ps   {\ensuremath{{\rm \,ps}}\xspace}

\newcommand{\stat}{\ensuremath{\mathrm{(stat)}}\xspace}
\newcommand{\syst}{\ensuremath{\mathrm{(syst)}}\xspace}

\newcommand{\chisq}{\ensuremath{\chi^2}\xspace}

\def\gsim{{~\raise.15em\hbox{$>$}\kern-.85em
          \lower.35em\hbox{$\sim$}~}\xspace}
\def\lsim{{~\raise.15em\hbox{$<$}\kern-.85em
          \lower.35em\hbox{$\sim$}~}\xspace}

\def\pt         {\mbox{$p_{\rm T}$}\xspace}

\newcommand{\lum} {\ensuremath{\mathcal{L}}\xspace}

\def\evtgen     {\mbox{\textsc{EvtGen}}\xspace}
\def\pythia     {\mbox{\textsc{Pythia}}\xspace}

\def\geant      {\mbox{\textsc{Geant4}}\xspace}

\def\photos     {\mbox{\textsc{Photos}}\xspace}

\def\tell1  {TELL1\xspace}
\def\ukl1   {UKL1\xspace}



%% file: title-LHCb-PAPER.tex

\begin{titlepage}
\pagenumbering{roman}

\vspace*{-1.5cm}
\centerline{\large EUROPEAN ORGANIZATION FOR NUCLEAR RESEARCH (CERN)}
\vspace*{1.5cm}
\hspace*{-0.5cm}
\begin{tabular*}{\linewidth}{lc@{\extracolsep{\fill}}r}
\ifthenelse{\boolean{pdflatex}}
{\vspace*{-2.7cm}\mbox{\!\!\!\includegraphics[width=.14\textwidth]{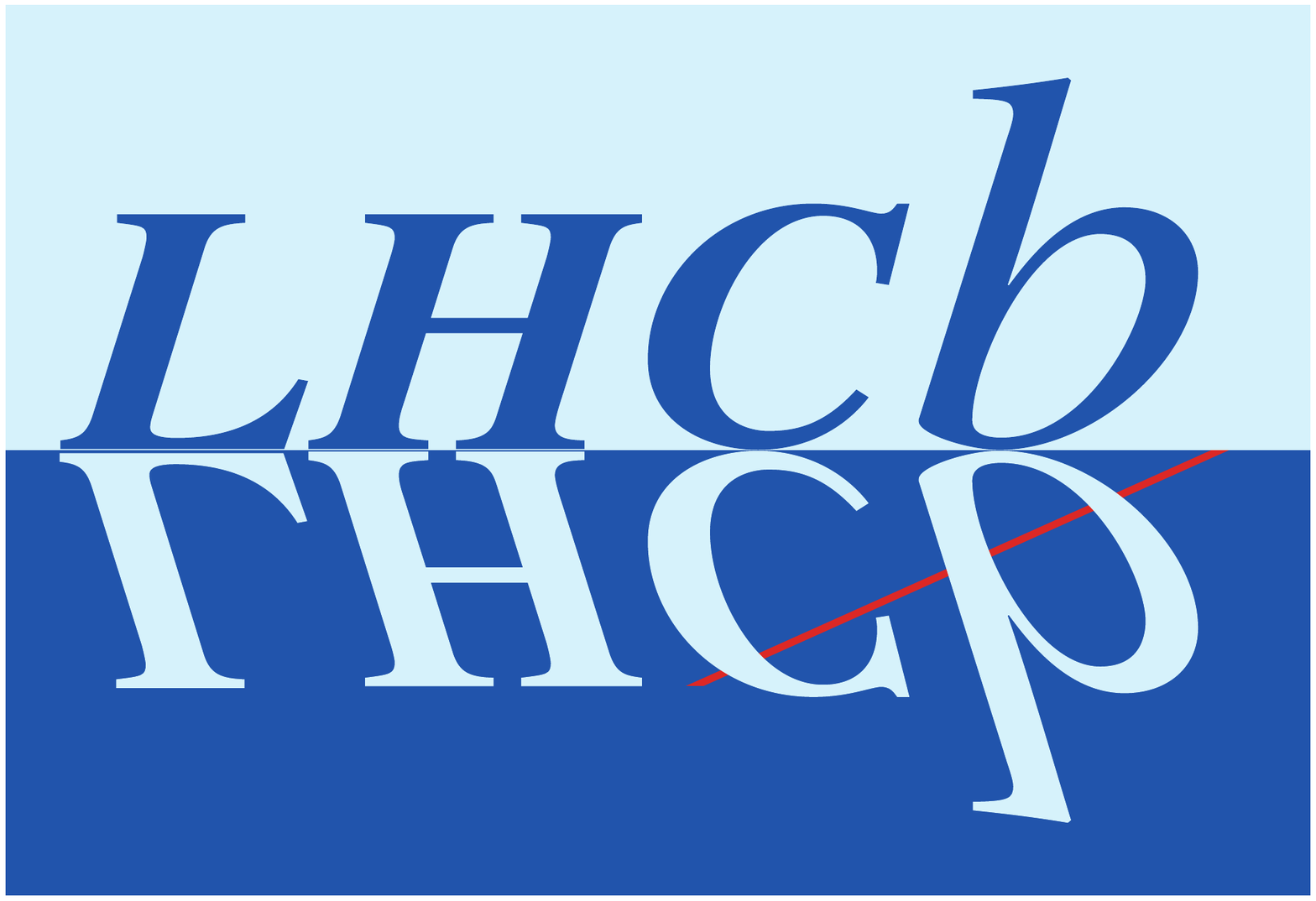}} & &}%
{\vspace*{-1.2cm}\mbox{\!\!\!\includegraphics[width=.12\textwidth]{lhcb-logo.eps}} & &}%
\\
 & & CERN-PH-EP-2012-349 \\  
 & & LHCb-PAPER-2012-039 \\  
 & & 10 January 2013 \\
 & & \\

\end{tabular*}

\vspace*{3.0cm}

{\bf\boldmath\huge
\begin{center}
Measurement of  \jpsi production in $pp$ collisions at $\sqrt{s}=2.76\tev$
\end{center}
}

\vspace*{1.5cm}

\begin{center}
The \lhcb collaboration\footnote{Authors are listed on the following pages.}
\end{center}

\vspace{\fill}

\begin{abstract}
  \noindent
The production of \jpsi mesons is studied with the LHCb detector using  data from $pp$ collisions at  $\sqrt{s}=2.76\tev$ corresponding to an integrated luminosity of $71\invnb$. The  differential cross-section for inclusive \jpsi production is measured as a function of its transverse momentum \pt.  The cross-section in the fiducial region $0<\pt<12\gevc$  and  
rapidity $2.0<y<4.5$ is measured to be $5.6\pm 0.1\,\stat\pm 0.4\,\syst\mub$, with the assumption of unpolarised \jpsi production. 
The fraction of \jpsi production from \bquark-hadron decays is measured to be $(7.1\pm 0.6\,\stat\pm0.7\,\syst)\%$.
\end{abstract}
\vspace*{2.0cm}

\begin{center}
  Submitted to JHEP
\end{center}

\vspace{\fill}

{\footnotesize
\centerline{\copyright~CERN on behalf of the \lhcb collaboration, license \href{http://creativecommons.org/licenses/by/3.0/}{CC-BY-3.0}.}}
\vspace*{2mm}

\end{titlepage}

\newpage
\setcounter{page}{2}
\mbox{~}
\newpage

\input{LHCb_authorlist.tex}

\cleardoublepage

%% file: LHCb_authorlist.tex
\centerline{\large\bf LHCb collaboration}
\begin{flushleft}
\small
R.~Aaij$^{38}$, 
C.~Abellan~Beteta$^{33,n}$, 
A.~Adametz$^{11}$, 
B.~Adeva$^{34}$, 
M.~Adinolfi$^{43}$, 
C.~Adrover$^{6}$, 
A.~Affolder$^{49}$, 
Z.~Ajaltouni$^{5}$, 
J.~Albrecht$^{35}$, 
F.~Alessio$^{35}$
% ...

{\footnotesize \it
% ...
$ ^{5}$Clermont Universit\'{e}, Universit\'{e} Blaise Pascal, CNRS/IN2P3, LPC, Clermont-Ferrand, France\\
$ ^{6}$CPPM, Aix-Marseille Universit\'{e}, CNRS/IN2P3, Marseille, France\\
% ...
$ ^{11}$Physikalisches Institut, Ruprecht-Karls-Universit\"{a}t Heidelberg, Heidelberg, Germany\\
% ...
$ ^{33}$Universitat de Barcelona, Barcelona, Spain\\
$ ^{34}$Universidad de Santiago de Compostela, Santiago de Compostela, Spain\\
$ ^{35}$European Organization for Nuclear Research (CERN), Geneva, Switzerland\\
% ...
$ ^{38}$Nikhef National Institute for Subatomic Physics, Amsterdam, The Netherlands\\
% ...
$ ^{43}$H.H. Wills Physics Laboratory, University of Bristol, Bristol, United Kingdom\\
% ...
$ ^{49}$Oliver Lodge Laboratory, University of Liverpool, Liverpool, United Kingdom\\
% ...
$ ^{n}$LIFAELS, La Salle, Universitat Ramon Llull, Barcelona, Spain\\
% ...
}
\end{flushleft}

%% file: paper_v6.tex
\newcommand{\prompt}{\ensuremath{\mathrm{prompt}~\jpsi}}
\newcommand{\fromb}{\ensuremath{\jpsi~\mathrm{from}~\bquark}}
\newcommand{\tpm}{$ & $\,\pm\,$ & $}

\section{Introduction}
This article presents the measurements of the differential inclusive  \jpsi production cross-section as a function of the \jpsi transverse 
momentum, and of the fraction of \jpsi mesons coming from the decay of a \bquark-hadron in $pp$ collisions at a centre-of-mass energy of 2.76\tev. 
The study is based on a sample corresponding to an integrated luminosity of $71\invnb$ collected in March 2011 with an average of one visible $pp$ interaction per recorded event. The main goal of 
this short run was to provide a reference for the study of Pb-Pb interactions carried out at the same centre-of-mass energy per nucleon-nucleon collision.  

Studies of \jpsi production have been performed by the LHC  experiments  using data taken at $\sqrt{s}=7\tev$\cite{LHCb-PAPER-2011-003,Aamodt:2011gj,Aad:2011sp,Chatrchyan:2011kc} as well as at lower energies\cite{Brambilla:2010cs}.
The data at $\sqrt{s}=2.76\tev$
provide an extra measurement to test theoretical models of \jpsi production in hadron collisions and are also used to obtain a measurement of \jpsi 
production from $b$-hadron decays.

The \lhcb detector~\cite{Alves:2008zz} is a single-arm forward spectrometer covering the \mbox{pseudorapidity} range $2<\eta <5$, designed 
for the study of particles containing \bquark or \cquark quarks. The detector includes a high precision tracking system consisting of a silicon-strip 
vertex detector (\velo) surrounding the $pp$ interaction region, a large-area silicon-strip detector located upstream of a dipole magnet with a bending 
power of about $4{\rm\,Tm}$, and three stations of silicon-strip detectors and straw drift tubes placed downstream. 
Charged hadrons are identified using two ring-imaging Cherenkov detectors. Photon, electron and hadron candidates are identified by a calorimeter system consisting of scintillating-pad and preshower detectors, an electromagnetic
calorimeter and a hadronic calorimeter. Muons are identified by a system which consists of five stations 
of alternating layers of iron and multiwire proportional chambers, with the exception of the centre of the first station, which uses triple-GEM detectors.

For the data used in this analysis, the \velo, which consists of two retractable halves surrounding the interaction region, was positioned during collisions with its sensitive area at a minimum distance  of  $13\,{\rm mm}$ from the beam instead of the nominal $8\,{\rm mm}$. This was necessary  to provide a larger aperture for the beam at the lower centre-of-mass energy of 2.76\tev.

The trigger\cite{Aaij:2012me} consists of a hardware stage, based
on information from the calorimeter and muon systems, followed by a
software stage, which applies a full event reconstruction.
Only the triggers used in this analysis are described here. At the hardware trigger level, a single muon candidate with \pt larger than 
$0.8\gevc$ is required.   
In the first stage of the software trigger a simplified event reconstruction is applied and one requires a \mumu candidate with invariant mass greater 
than 2.7\gevcc.  In the second stage a full event reconstruction is performed  and only events with a \mumu  pair with invariant mass 
within 120\mevcc of the known \jpsi mass~\cite{Nakamura:2010zzi} are retained.  

\section{Event selection}

The analysis strategy is based upon that described in Ref.~\cite{LHCb-PAPER-2011-003}. Candidate \jpsi mesons are formed from pairs of opposite-sign charged particles reconstructed in the fiducial region $2<\eta<5$ by the full tracking system using algorithms adapted to the \velo at its displaced position. Each particle must have \pt above 0.7\gevc\ and be identified as a muon. The two muons are required to originate from a common vertex, and only 
candidates with a \chisq probability of the vertex fit larger than 0.5\% are kept. Events are selected in which at least one primary vertex is 
reconstructed from at least three \velo tracks, excluding the two signal muon tracks from the \jpsi decay. A \velo track is required to have at least three hits on a straight line in the radial strips of the  detector. 

The Monte Carlo samples used for this analysis are based on the \pythia 6.4 generator~\cite{Sjostrand:2006za} configured with the parameters detailed in 
Ref.~\cite{LHCb-PROC-2010-056}. 
The \evtgen package~\cite{Lange:2001uf} is used to generate hadron decays, in particular for \jpsi and \bquark-hadrons.
The interaction of the generated particles with the detector and its
response are implemented using the \geant toolkit~\cite{Allison:2006ve, *Agostinelli:2002hh} as described in
Ref.~\cite{LHCb-PROC-2011-006}.
Radiative corrections to the decay $\jpsi \to \mumu$ are generated using 
the \photos package~\cite{Golonka:2005pn}. The simulated position of the \velo corresponds to that in the data.

\section{Cross-section determination}

The  differential cross-section for \jpsi production in a \pt bin is given by
\begin{equation}
\frac{{\rm d}\sigma}{{\rm d}\pt} = \frac{N\left(\jpsi\to\mumu\right)}{\lum \times \epsilon_{\rm tot}
\times {\cal B}\left(\jpsi\to\mumu\right)\times  \Delta \pt}\label{eq:sigma}\,,
\end{equation}
where $N\left(\jpsi \to \mumu\right)$ is the number of observed $\jpsi \to \mumu$ signal decays in the given bin, $\epsilon_{\rm tot}$ the \jpsi detection efficiency per \pt bin
(including both acceptance and trigger), \lum the integrated luminosity, 
${\cal B}\left(\jpsi\to\mumu\right)=(5.93\pm0.06)\times 10^{-2}$~\cite{Nakamura:2010zzi} the branching fraction of the $\jpsi \to \mumu$ decay,
and $\Delta \pt$ the \pt bin size.

\begin{figure}[!tb]
\centering
\includegraphics[width=16cm]{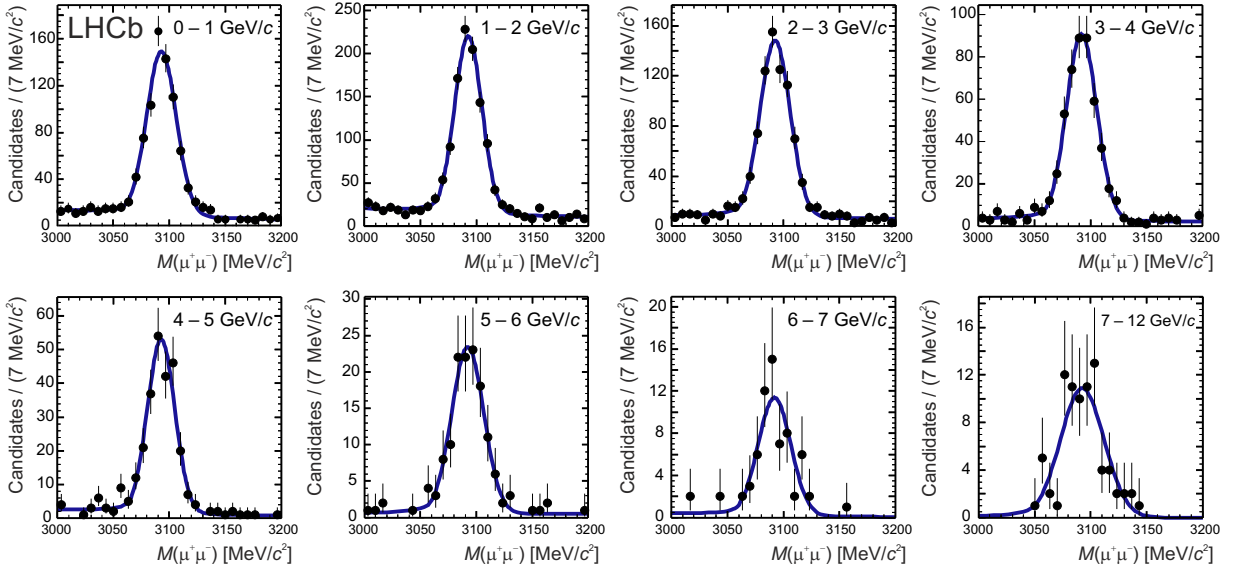}
\caption{\small Dimuon mass distributions, with fit results superimposed, in bins of  \pt. Results for $\pt>7\gevc$ are merged in the last bin. } 
\label{fig:Fit_pt}
\end{figure}

The number of signal \jpsi  mesons per \pt  bin is determined from an extended unbinned maximum likelihood fit to the invariant mass distribution of 
the reconstructed \jpsi candidates in the interval $3.0<M_{\mu\mu}<3.2\gevcc$, where the signal is described by a Crystal Ball 
function~\cite{Skwarnicki:1986xj} and the combinatorial background by an exponential distribution. 
Figure~\ref{fig:Fit_pt} shows the \jpsi invariant mass distribution together with the fit results for each \pt  bin, where results for $7<\pt<12\gevc$ are merged in the 
last bin. 

There are two main sources that contribute to the inclusive \jpsi  sample. Those produced at the $pp$ collision point,
either directly or from the decay of a directly produced higher mass charmonium state, are called \prompt. The second source, \fromb, are those produced in the
decay of a $b$-hadron. Their production is displaced from the $pp$ collision point because of the relatively large
\bquark lifetime. The two sources are statistically separated using the measured \jpsi pseudo-decaytime, defined as
\begin{equation}\label{eq:tz}
t_z = \frac{(z_{\jpsi}-z_{\rm PV}) \times M_{\jpsi}}{p_z}\,,
\end{equation}
where $z_{\jpsi}$ and $z_{\rm PV}$ are the positions along the beam axis of the \jpsi decay vertex and of the primary vertex
refitted after removing the two muon tracks from the \jpsi candidate; $p_z$ is the measured \jpsi momentum in the beam direction and $M_{\jpsi}$ the known \jpsi mass~\cite{Nakamura:2010zzi}. Given that \bquark-hadrons are not fully reconstructed, the \jpsi momentum is used instead of the exact  \bquark-hadron 
momentum and the $t_z$ variable provides a good estimate of  the \bquark-hadron decaytime. 

The fraction of \fromb\ is determined from a simultaneous fit to the total pseudo-decaytime $t_z$ and $\mumu$ invariant mass. Due to the small number of \jpsi candidates, the fraction of \fromb\ is  computed over the full \pt interval from 0 to 12 \gevc.
The signal decaytime distribution is described by a delta function at $t_z=0$ for the prompt \jpsi component and an exponential decay function for the \fromb\ component. The function describing the $t_z$ distribution of the signal is therefore
\begin{equation}
f_{\rm signal} (t_z;f_{\rm p},f_{\rm b},\tau_{\rm b}) = f_{\rm p}\,\delta(t_z)+\theta(t_z)f_{\rm b}\frac{ e^{-\frac{t_z}{\tau_{\rm b}}}}{ \tau_{\rm b}}\,,
\end{equation}
where $\theta(t_z)$ is the step function, $f_{\rm p}$ and $f_{\rm b}$ are the fractions of prompt \jpsi and \fromb\  in the sample, and $\tau_{\rm b}$ the \bquark-hadron 
pseudo-lifetime. In the fit, $\tau_b$ is fixed to the value of 1.52\ps, as obtained from simulation. 
The prompt and \bquark components of the signal function are convolved with a triple-Gaussian resolution function
\begin{equation}
f_{\rm res}(t_z;\mu,\sigma_1,\sigma_2, \sigma_3 , \beta , \beta') = 
\frac{\beta}{\sqrt{2\pi}\sigma_1}\,e^{-\frac{(t_z-\mu)^2}{2\sigma_1^2}}+ 
\frac{\beta'}{\sqrt{2\pi}\sigma_2}\,e^{-\frac{(t_z-\mu)^2}{2\sigma_2^2}}+
\frac{1-\beta-\beta'}{\sqrt{2\pi}\sigma_3}\,e^{-\frac{(t_z-\mu)^2}{2\sigma_3^2}}.
\end{equation}
The parameter $\mu$ is the bias of the $t_z$ measurement, and $\beta$ and $\beta'$ the fractions of the first two Gaussian functions.
The background consists of random combinations of muons from semi-leptonic \bquark and \cquark decays, 
which tend to produce positive $t_z$ values, as well as of mis-reconstructed tracks from decays in flight of kaons and pions, which contribute both 
to positive and negative $t_z$ values. The background $t_z$ distribution is parameterised with an empirical  function based on the shape obtained from the \jpsi mass sidebands. It is taken as the sum of a  delta function and three exponential components, two for positive $t_z$ and 
one for negative $t_z$. The exponential parameter, $\tau_{\rm L}$,  is common to the larger positive and negative lifetime exponential components. The explicit form is
\begin{equation} 
\label{backfunc}
f_{\rm bckg}(t_z) = \left(1-f_1-f_{\rm L}\right)\delta(t_z) + \theta(t_z)  f_1\, \frac{e^{-\frac{t_z}{\tau_1}}}{\tau_1} +  
f_{\rm L}\, \frac{e^{-\frac{|t_z|}{\tau_{\rm L}}}}{2 \tau_{\rm L}},
\end{equation}
and is convolved with the same resolution function $f_{\rm res}$ as the signal.

\begin{figure}[!tb]
\centering
\includegraphics[width=7.9cm]{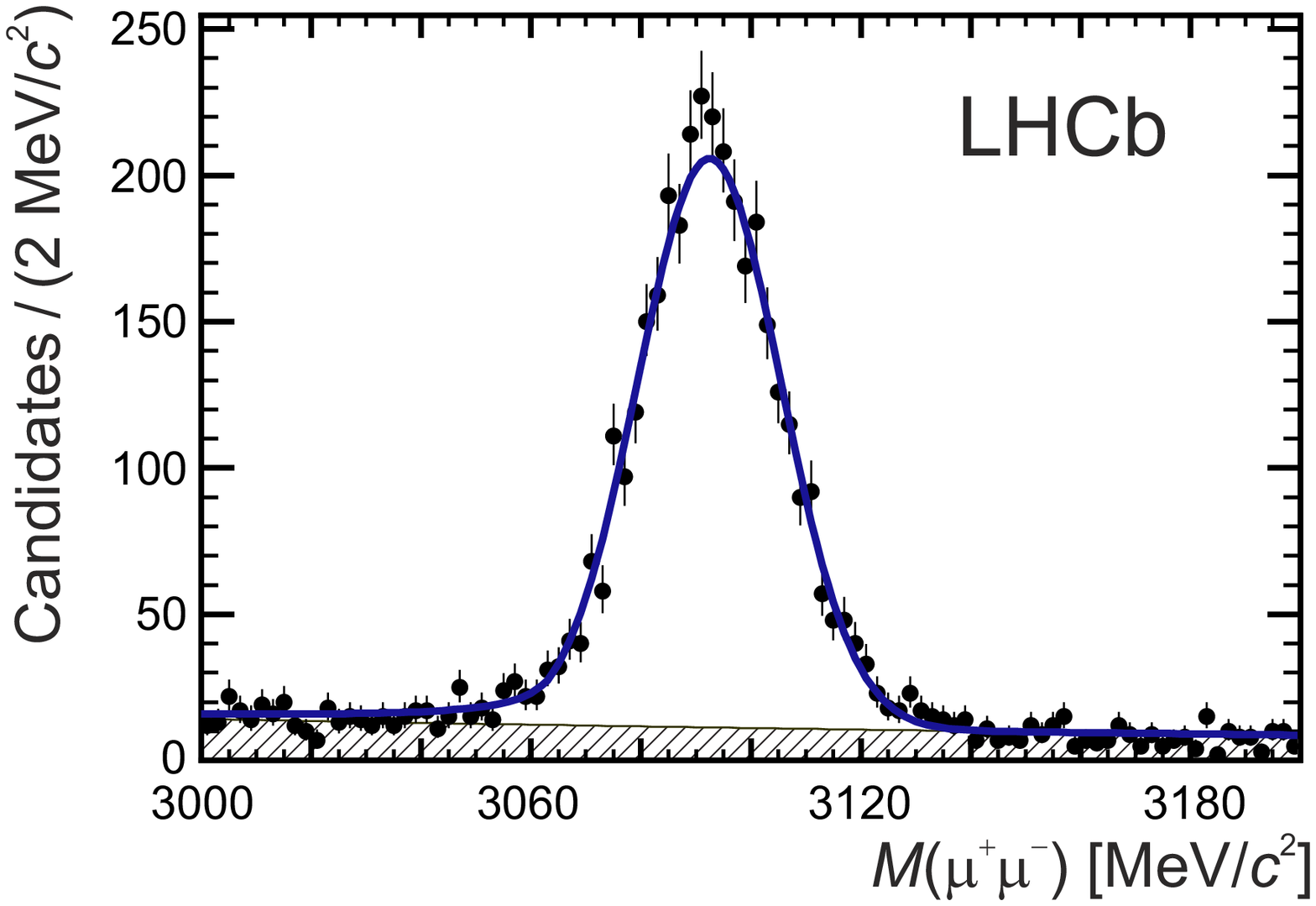}
\includegraphics[width=7.9cm]{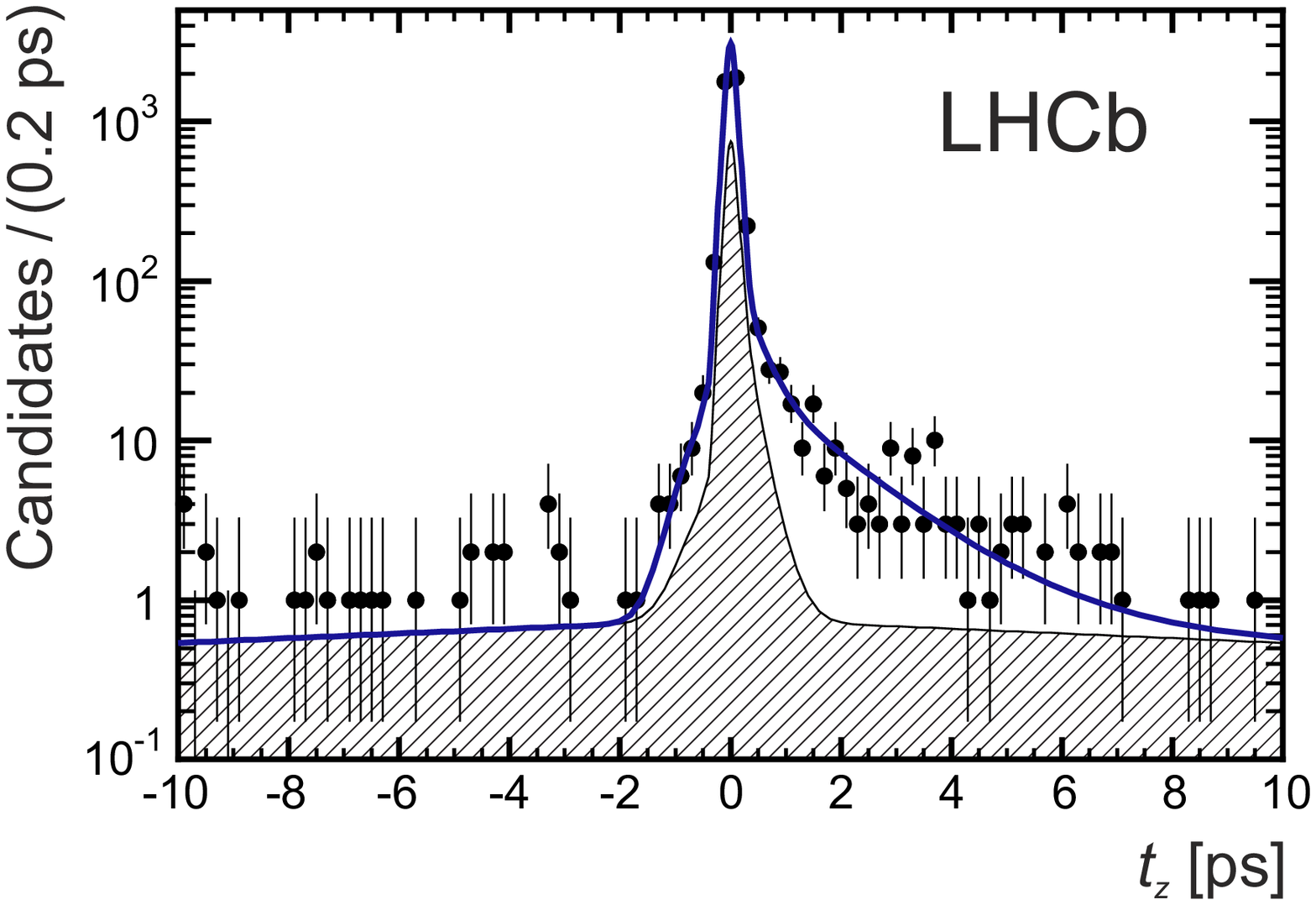}
\caption{\small Distributions of the data with projections of the fit superimposed for (left) the dimuon invariant mass and (right) $t_z$.  
The thick blue line is the total fit function and the hatched area the background component.}
\label{fig:Fit_mass&tz}
\end{figure}

The function used to describe the $t_z$ distribution is therefore
\begin{equation}\label{eq:tzfit}
\begin{split}
f(t_z  ; f_{\rm p} , f_{\rm b} , \mu , \sigma_1 , \sigma_2 , \sigma_3 , \beta , \beta' ,\tau_{\rm b} )  = 
&\left(f_{\rm p}\, \delta(t_z)+f_{\rm b} \, \frac{e^{-\frac{t_z}{\tau_{\rm b}}}}{\tau_{\rm b}}+ \left(1-f_{\rm p}-f_{\rm b} \right) f_{\rm bckg}( t_z )\right) \otimes \\
& f_{\rm res}(t_z;\mu,\sigma_1,\sigma_2, \sigma_3,\beta, \beta') \,,
 \end{split}
\end{equation}
where all parameters except $\tau_{\rm b}$ are freely varied. 
The total fit function is the sum of the products of the mass and $t_z$ fit functions for the signal and background.  
Figure~\ref{fig:Fit_mass&tz} shows the distributions of the dimuon invariant mass and $t_z$ with the projections of the fit superimposed. 
The invariant mass resolution is $13.0\pm0.3\mevcc$.  The parameter $\mu$ describing the bias of the $t_z$ resolution function is $2.3\pm2.0$~fs and the RMS of the $t_z$ resolution function is 84~fs. As a measure of the fit quality, a $\chi^2$ is calculated using a binned event distribution. The resulting fit probability for the $t_z$ distribution is 90\%.
The fit gives a total yield of $3399\pm65$ \jpsi signal decays. 

The fraction of signal \jpsi coming from \bquark-hadron decays is measured to be $ F_{\rm b}=\frac{f_b}{f_p+f_b}=(6.7\pm0.6)\%$. 
An absolute correction of $0.4\%$ is applied based on simulation to take into account a bias produced by events in which \bquark-hadron 
decay products, other than the muons from the \jpsi, are wrongly used to reconstruct the primary vertex. This leads to the result 
$F_{\rm b}=(7.1\pm0.6)\%$ where the uncertainty is only statistical. 

\begin{figure}[!tb]
\centering
\includegraphics[width=9.0cm]{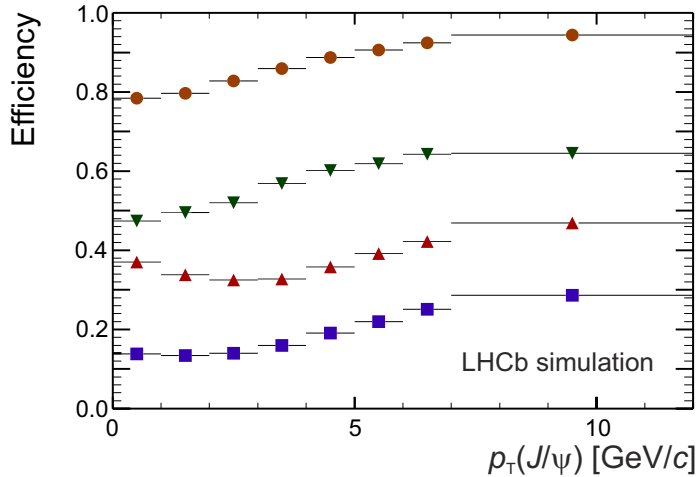}
\caption{\small Acceptance (orange circles), reconstruction (upward-facing red triangles), trigger (downward-facing green triangles) and total (blue squares) \jpsi efficiency, as a function of \pt. The efficiencies are evaluated from a Monte Carlo simulation in which the \jpsi  is produced unpolarised. }
\label{TotalEfficiency}
\end{figure}

A simulated sample of inclusive, unpolarised \jpsi mesons is used to estimate the geometrical acceptance in each \pt bin. The reconstruction 
efficiency, which combines the \jpsi meson detection, reconstruction and selection efficiencies, is also computed from simulation as a 
function of \pt  and is corrected to account for the difference observed  in the tracking efficiency between data and simulation at 
$\sqrt{s}=7\tev$. This correction is about 1\%.  The efficiency of the hardware trigger is determined directly from data using 
 a large inclusive \jpsi sample at $\sqrt{s}=7\tev$ triggered and selected with the same requirements as those used in this analysis: the efficiency is
 calculated in small bins of the \jpsi transverse momentum and rapidity and weighted according to the \pt and $y$ distributions as given 
 by the simulation at 2.76\tev.
 The efficiency of the software trigger, which 
makes use of the \velo information, is determined from simulation since the data  at $\sqrt{s}=7\tev$ were taken with the \velo in the closed position. 
The total efficiency, calculated as the product of acceptance, reconstruction and trigger efficiencies, and its components are displayed in Fig.~\ref{TotalEfficiency} as a 
function of \pt. A non-zero polarisation of the \jpsi at production can affect the total efficiency~\cite{LHCb-PAPER-2011-003}. The results quoted in this article assume that the \jpsi mesons are produced unpolarised. 

\section{Luminosity determination}

To determine the integrated luminosity, an effective interaction rate is continuously measured during data taking and an absolute calibration is 
performed with a dedicated van der Meer (VDM) scan~\cite{vdm}. The strategy is similar to that developed for the $\sqrt{s}=7\tev$ running~\cite{LHCb-PAPER-2011-015}.

\renewcommand{\arraystretch}{1.2}
\begin{table}[!b]
\tabcolsep 4mm
\begin{center}
\caption{\small \label{lumisigma}Relative systematic uncertainties on the luminosity (\%).}
\begin{tabular}{@{}lll@{}}
\toprule
\multicolumn{3}{l}{Uncertainty on relative normalisation} \\
& Counter stability & \phantom{7}0.5 \\
& $\mu$ variation among bunch crossings & \phantom{7}0.5 \\ \midrule
\multicolumn{3}{l}{Uncertainty on absolute normalisation} \\
& Statistical error of the VDM scan& \phantom{7}0.2 \\
& Total beam current & \phantom{7}2.7 \\
& Individual bunch population & \phantom{7}0.9 \\
& Protons outside nominal bunches & \phantom{7}0.4\\
& Length scale calibration & \phantom{7}1.0\\
& Non-reproducibility in similar scans at $\sqrt{s}=7\tev$ & \phantom{7}2.1\\ \midrule
& Total uncertainty & \phantom{7}3.8 \\ 
\bottomrule
\end{tabular}
\end{center}
\end{table}

The VDM method exploits the ability to move the beams  in both transverse coordinates  with high precision and thus to scan the colliding beams 
with respect to each other. 
 The limiting systematic uncertainty affecting the VDM measurement arises from the knowledge of the number of protons in the  colliding bunch pairs. These are
measured with two types of beam current transformers installed in the LHC~\cite{bct1,*bct2,*bct3}. 
The DCCT (DC Current Transformer) measures the total beam
current, and is thus used to constrain the total number of particles. The uncertainty associated with  the DCCT calibration is 2.7\%~\cite{bcnwg1,*bcnwg2,*Ohm:2009pf}. The other
transformer, the FBCT (Fast Beam Current Transformer) is used to measure the relative charges of the individual bunches. The uncertainty in its 
offset and linearity contributes a 0.9\%  uncertainty to the overall luminosity~\cite{bcnwg1,*bcnwg2,*Ohm:2009pf}.  A small fraction of protons in the LHC may be captured 
outside the nominally filled bunch slots. This contribution, which needs to be subtracted from the DCCT measurement, is estimated to be 2.5\% from 
the number of beam-gas events in nominally empty bunch crossings. Due to the small number of such events and uncertainties in the trigger 
efficiency, the subtraction introduces a cross-section uncertainty of 0.4\%. The uncertainty in  the length-scale calibration, which affects the beam 
separation values, contributes 1\% to the systematic uncertainty in the luminosity. Finally, a 2.1\% uncertainty is assigned to account for a non-reproducibility 
of the VDM results observed when performing similar luminosity calibration measurements at $\sqrt{s}=7\tev$, as decribed in Ref.~\cite{LHCb-PAPER-2011-015}.

The integrated luminosity for the runs considered in this analysis is measured to be $70.6\pm2.7\invnb$. A summary of the contributions to the overall luminosity uncertainty is provided in Table~\ref{lumisigma}.  The uncertainties are
uncorrelated and therefore added in quadrature.

\section{Systematic uncertainties}\label{sec:systematics}

The different contributions to the systematic uncertainty affecting the cross-section measurement are summarised in Table~\ref{syst}.
Correction factors estimated directly from data to take into account residual differences between simulation and data
are also detailed.

The influence of the choice of the fit function used to describe the shape of the dimuon mass distribution  is estimated by fitting the \jpsi invariant 
mass distribution with the sum of two Crystal Ball functions. The relative difference of 2.2\% in the number of signal events is taken as systematic 
uncertainty.

A fraction of \jpsi events have a lower mass because of the radiative tail. Based on Monte Carlo studies,  5\% of the \jpsi signal is estimated to be outside 
the analysis mass window ($M_{\mu\mu}<3.0\gevcc$) and not counted as signal. The fitted signal yields are therefore corrected, and an 
uncertainty of 1\% is assigned to the cross-section measurement based on a comparison between the radiative tail observed in data and simulation.

To cross-check and assign a systematic uncertainty to the Monte Carlo determination of the muon identification efficiency,  the single track muon 
identification efficiency is measured on data using a tag-and-probe method. This method reconstructs \jpsi candidates in which one muon is identified by 
the muon system (``tag'') and the other one (``probe'') is identified by selecting a track with a  minimum-ionising energy deposition in the 
calorimeters. The absolute muon identification efficiency is then evaluated on the probe muon, as a function of the muon momentum and found to be larger than 95\%. The ratio of the 
muon identification efficiency measured in data to that obtained in the simulation is convolved with the momentum distribution of muons from 
\jpsi to obtain an efficiency correction. This factor is found to be $1.024\pm0.011$ and is consistent with being 
constant over the full \jpsi transverse momentum and rapidity range; the error on the correction factor is included as a systematic uncertainty.

\renewcommand{\arraystretch}{1.2}
\begin{table}[!t]
\tabcolsep 4mm
\begin{center}
\caption{\small \label{syst}Relative systematic uncertainties on the cross-section results and on the fraction
of \jpsi mesons from $b$-hadron decay (\%).}
\begin{tabular}{@{}lll@{}}
\toprule
& Source & Systematic uncertainty \\\midrule
\multicolumn{3}{l}{Correlated between bins} \\
& Mass fits & $\phantom{1}2.2$ \\
& Radiative tail & $\phantom{1}1.0$ \\
& Muon identification & $\phantom{1}1.1$ \\
& Tracking efficiency & $\phantom{1}0.8\ {\rm to}\ 1.1$ \\
& Track \chisq & $\phantom{1}2.0$ \\
& Vertexing & $\phantom{1}0.3$ \\
& Model dependence & $\phantom{1}4.5$ \\
& ${\cal B}(\jpsi\to\mumu)$ & $\phantom{1}1.0$ \\
& Luminosity & $\phantom{1}3.8$ \\
\midrule
\multicolumn{3}{l}{Uncorrelated between bins} \\
& Trigger  & $\phantom{1}1.6\ {\rm to}\ 7.7$ \\
\midrule
\multicolumn{3}{l}{Applied only to \jpsi from \bquark fraction} \\
& $t_z$ fit & $10.0$ \\
\midrule
\multicolumn{3}{l}{Applied only to $\sigma(pp \to \bquark\bquarkbar X)$} \\
& ${\cal B}(b\to\jpsi X)$ & $\phantom{1}8.6$ \\
\bottomrule
\end{tabular}
\end{center}
\end{table}

Studies at $\sqrt{s}=7\tev$ have shown that the Monte Carlo simulation reproduces the determination from data of  the efficiency to reconstruct the two muon tracks from the \jpsi 
decay within 0.8\% to 1.1\%, depending on the \jpsi transverse momentum. This difference is taken as a systematic uncertainty.
An additional uncertainty of 1\% per track is assigned to cover differences in the efficiency of the track \chisq/ndf  cut between data and simulation. 
Similarly, for the selection based on the \jpsi vertex \chisq probability, a difference below 0.3\% is measured between the cut efficiency computed in data and 
simulation, which is assigned as systematic uncertainty. 
To take into account the model dependence of the simulation in the efficiency calculation, the main parameters of the \pythia 6.4 generator
related to prompt \jpsi production were varied. 
These parameters define the minimum \pt cut-offs for regularising the cross-section. A 4.5\% effect on the total efficiency was observed.

The hardware trigger efficiency is determined using a  sample of events at $\sqrt{s}=7\tev$ that would still be triggered if the \jpsi 
candidate were removed. The software trigger efficiency is obtained from the simulation. Its uncertainty  is evaluated by 
comparing true and measured trigger efficiency using a trigger-unbiased sample of simulated \jpsi events.

Uncertainties related to the $t_z$ fit procedure are taken into account by varying the slope of the exponential function of the \fromb\ component
by its uncertainty in the simulation (2\%). 
The  resulting 10\% variation of the number of \fromb\  is used as a systematic uncertainty that affects the measurement of $F_{\rm b}$. The influence of the background parametrisation was studied by varying the number of exponential functions in Eq.~\eqref{backfunc} and found to be negligible.
Furthermore, an uncertainty of 8.6\% on the average branching fraction of \bquark decays to a final state containing a \jpsi meson
contributes to the uncertainty on the extrapolation to the total $\bquark\bquarkbar$ cross-section.

\section{Results}

\begin{table}[!tb]
\caption{\small \label{dsigdpt} Differential cross-section ${\rm d}\sigma / {\rm d}\pt$ at $\sqrt{s}=2.76\tev$ for inclusive \jpsi production in bins of \pt. The rapidity range covered is $2.0<y<4.5$. The first uncertainty is 
statistical and the second is systematic.}
\begin{center}
\begin{tabular}{@{}r@{}c@{}lr@{}r@{}c@{}r@{}c@{}}
\toprule
\multicolumn{3}{@{}l}{$\pt\,({\rm GeV}/c)$} & \multicolumn{5}{r@{}}{${\rm d}\sigma / {\rm d}\pt$ [nb/(\gevc)]} \\
\midrule
$0$&$-$&$1$     & $1270                \tpm60                   \tpm 130\phantom{11111}$   \\
$1$&$-$&$2$     & $1780                \tpm70                   \tpm 160\phantom{11111}$   \\
$2$&$-$&$3$     & $1290                 \tpm 50                  \tpm \phantom{1}90\phantom{11111}$    \\
$3$&$-$&$4$     & $\phantom{1}\,700\tpm 40                  \tpm \phantom{1}50\phantom{11111}$      \\
$4$&$-$&$5$     & $\phantom{1}\,313\tpm 22                   \tpm \phantom{1}24\phantom{11111}$      \\
$5$&$-$&$6$     & $\phantom{1}\,142\tpm 13                   \tpm \phantom{1}10\phantom{11111}$        \\
$6$&$-$&$7$     & $\phantom{1\,1}61\tpm \phantom{1}8  \tpm \phantom{11}4\phantom{11111}$           \\
$7$&$-$&$12$    & $\phantom{1\,1}14\tpm \phantom{1}2   \tpm \phantom{11}1\phantom{11111}$           \\
\bottomrule
\end{tabular}
\end{center}
\end{table}

\begin{figure}[!t]
\centering
\includegraphics[width=11.0cm]{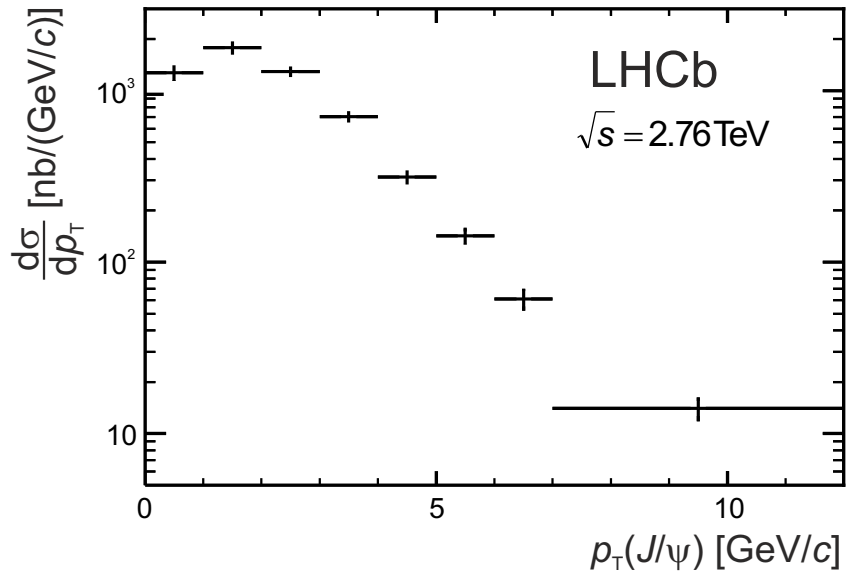}
\caption{\small Differential production cross-section for inclusive \jpsi production in the rapidity range $2.0<y<4.5$ as a function of \pt. The vertical error bars are the quadratic sums of the statistical and systematic uncertainties.} \label{sigmaresults_fromb}
\end{figure}

The measured  differential cross-section for inclusive \jpsi production as a function of  \pt,  after all corrections and assuming no polarisation, is 
given in Table~\ref{dsigdpt}  and displayed in Fig.~\ref{sigmaresults_fromb}. 
The integrated inclusive cross-section for \jpsi production in the defined fiducial region is
\begin{equation*}\label{sigma}
\sigma\left(\jpsi, \, \pt <12\,\gevc,\,2.0<y<4.5\right)  \, = \, 
5.6\pm 0.1\pm 0.4\mub.
\end{equation*}
The first uncertainty is statistical and the second systematic. Studies indicate that this result could change by up to 20\% assuming fully longitudinal or fully transverse \jpsi polarisation~\cite{LHCb-PAPER-2011-003}. 
The fraction of \fromb\ is measured to be
\begin{equation*}\label{fb}
F_b = (7.1 \pm 0.6 \pm 0.7)\%
\end{equation*}
in the same acceptance range, $\pt<12\gevc$ and $2.0<y<4.5$.

From the above results, one can deduce
\begin{equation*}
\sigma\left(\fromb, \, \pt <12\,\gevc,\,2.0<y<4.5\right)  \, = \, 
400\pm 35\pm 49\nb,
\end{equation*}
in good agreement with the theoretical prediction of $370^ {+ 170}_ {-110}\nb$, based on NLO calculations described in Ref.~\cite{Cacciari}.
In addition, the total $\bquark\bquarkbar$ production cross-section is computed as 
\begin{equation}
\sigma(pp \to b\overline{b} X) = 
\alpha_{4\pi} \,\frac{\sigma\left(\jpsi, \,\pt <12\,\gevc,\,2.0<y<4.5\right)\times F_{\rm b}}{2 \, {\cal B}(b\to\jpsi X)},
\end{equation}
where the factor $\alpha_{4\pi}=6.3$ is an extrapolation factor of the cross-section from the measured
to the full kinematic region. This factor is obtained using the simulation software described previously.
The inclusive $b\to\jpsi X$ branching fraction is \mbox{${\cal B}(b\to\jpsi X)=(1.16\pm0.10)\%$}~\cite{Nakamura:2010zzi}. 
The resulting total $b\overline{b}$ cross-section is \mbox{$\sigma(pp \to b\overline{b} X) = 110\pm 9 \pm 16 \, \upmu{\rm b}$}. 
No systematic uncertainty has been included for the extrapolation factor $\alpha_{4\pi}$ estimated from the simulation. The value of the
extrapolation factor given by NLO calculations is 6.1~\cite{Cacciari}.

\section{Conclusions}

The differential cross-section for inclusive \jpsi production is measured as a function of the \jpsi transverse momentum in the forward region, 
$2.0<y<4.5$. The analysis is based on a data sample corresponding to an integrated luminosity of $71\invnb$ collected 
by the \lhcb experiment at the Large Hadron Collider at 
a centre-of-mass energy of $\sqrt{s}=2.76\tev$. 
The results obtained are in good agreement with earlier measurements of the inclusive \jpsi production cross-section in $pp$ collisions at the
same centre-of-mass energy, performed by ALICE in the region $2.5<y<4.0$~\cite{Abelev:2012kr}. 
A first measurement of the production of \jpsi from $b$-hadron decays at 2.76\tev is also obtained.

%% file: acknowledgements.tex
\section*{Acknowledgements}

\noindent We thank M.~Cacciari for providing theoretical predictions of the $b\overline{b}$ production cross-section in the LHCb acceptance.
We express our gratitude to our colleagues in the CERN
accelerator departments for the excellent performance of the LHC. We
thank the technical and administrative staff at the LHCb
institutes. We acknowledge support from CERN and from the national
agencies: CAPES, CNPq, FAPERJ and FINEP (Brazil); NSFC (China);
CNRS/IN2P3 and Region Auvergne (France); BMBF, DFG, HGF and MPG
(Germany); SFI (Ireland); INFN (Italy); FOM and NWO (The Netherlands);
SCSR (Poland); ANCS/IFA (Romania); MinES, Rosatom, RFBR and NRC
``Kurchatov Institute'' (Russia); MinECo, XuntaGal and GENCAT (Spain);
SNSF and SER (Switzerland); NAS Ukraine (Ukraine); STFC (United
Kingdom); NSF (USA). We also acknowledge the support received from the
ERC under FP7. The Tier1 computing centres are supported by IN2P3
(France), KIT and BMBF (Germany), INFN (Italy), NWO and SURF (The
Netherlands), PIC (Spain), GridPP (United Kingdom). We are thankful
for the computing resources put at our disposal by Yandex LLC
(Russia), as well as to the communities behind the multiple open
source software packages that we depend on.